\documentstyle[psfig,epsfig,epsf]{ta}

\def\be{\begin{equation}}
\def\ee{\end{equation}}
\def\bea{\begin{eqnarray}}
\def\eea{\end{eqnarray}}

\begin{document}

\title{Removing Line Interference from Gravitational Wave
Interferometer Data}

\author{Alicia M.\ SINTES and Bernard F.\ SCHUTZ\\
{\it Max-Planck-Institut f\"ur Gravitationsphysik (Albert-Einstein-Institut)
  Am M\"uhlenberg 1. 14476 Golm. Germany, 
  sintes@aei-potsdam.mpg.de}}

\maketitle

\section*{Abstract}

We describe a procedure to identify and remove a class of 
interference lines from gravitational wave interferometer data.
We illustrate the usefulness of this technique applying it
to prototype interferometer data and removing all those lines
corresponding to the external electricity main supply and related features.


\section{Introduction}

In the measured noise spectrum of the different  gravitational wave 
interferometer prototypes, we observe peaks due 
external interference [1,2].
 The most numerous are powerline frequency harmonics.
We have shown how to model and remove these lines very
 effectively
using a technique we call coherent line removal ({\sc clr}) [3].
In addition to those lines appearing at multiples of 50 (or 60) Hz,
there are other interference lines whose frequencies change in step
with the  supply frequency, but not at the harmonic frequencies.
The easiest way to detect their presence is by studying in detail the 
spectrogram
(see figure 1). These lines, although they are not as
powerful as the harmonics,  are spread over the whole spectrum.
From a data analysis point of view, we try to  develop a technique 
able to remove this interference while producing a minimum 
disturbance to the underlying noise background [4].


\begin{figure} [t]
\begin{center}
\centerline{\vbox{ 
\psfig{figure=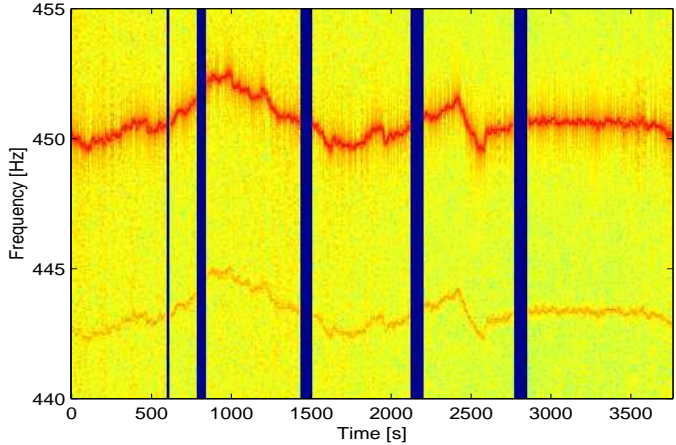,height=6.0cm,width=9.0cm} 
}} 
\end{center}
\caption[]{Zoom of the spectrogram of the prototype data. 
The line near 450 Hz  corresponds to the 9th harmonic of the
external electricity supply. The weaker line near 443 Hz is
one example of the many lines present in the spectrum that have a
similar time-frequency evolution to that of the harmonics of 50 Hz.
} 
\end{figure} 

In this paper, we illustrate the usefulness of these techniques by applying
them to the data produced by the Glasgow laser interferometer in March 1996.
 As a result the interference is attenuated or eliminated
by cancellation in the time domain and the power spectrum appears 
completely clean allowing the detection of signals that were 
buried in the
interference. Therefore, these new methods appear to be good news as far
as searching for continuous waves  is concerned.
 The removal 
 reduces the level of non-Gaussian noise,  improving
  the sensitivity of the detector to short
bursts of gravitational waves as well.

\section{Line removal}

{\sc clr} is an algorithm able to remove interference present in the
 data while preserving the stochastic detector noise.  {\sc clr}
 works when the interference is present in many harmonics, as long as
   they remain 
 coherent with one another. It can remove the external
 interference without removing any \lq single line' signal buried by the
 harmonics. The algorithm works even when the interference frequency changes.
{\sc clr} can be used to remove all harmonics of periodic or 
broad-band signals (e.g., those which change frequency in time), even when 
there is no external reference source.   It
 assumes that the interference has the form
\be
y(t)=\sum_n a_n m(t)^n + \left( a_n m(t)^n\right)^* \ ,
\label{e3}
\ee
where $a_n$ are
complex amplitudes and  $m(t)$ is a nearly monochromatic function
 near
a frequency $f_0$.
The idea is to
 use the information in the different harmonics of the interference 
to construct a function $M(t)$ that is  as close a replica as possible of
$m(t)$ and then construct a function close to $y(t)$ which 
is subtracted from the output  of the system cancelling the interference.
The key is that real gravitational wave signals will not be present
with multiple harmonics and that $M(t)$ is constructed from many
frequency bands with independent noise. Hence, {\sc clr} will little
affect the statistics of the noise in any one band and
any gravitational wave signal masked by the interference can
be recovered without any disturbance.

%
\begin{figure}[h!]
\begin{center}
\centerline{\vbox{ 
\psfig{figure=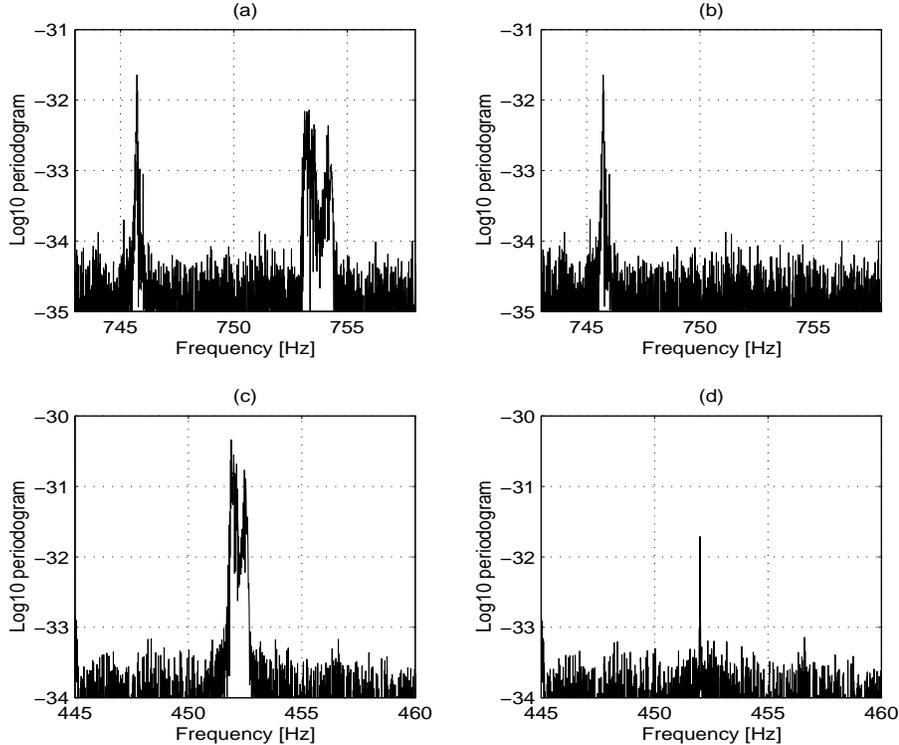,height=10.0cm,width=12.0cm} 
}} 
\end{center}
\caption[]{Decimal logarithm of the
periodogram of $2^{19}$ points of the prototype data. 
(a) One of the harmonics near 754 Hz.
(b) The same data after the removal of the interference using
{\sc clr}.
(c)  The same experimental  data
with an artificial  signal added at 452 Hz. 
(d) The data in (c) after the removal of the interference, revealing that
the signal remains detectable. 
} 
\end{figure} 
In figure 2, we show  the performance of {\sc clr}  on two minutes of data.
We can see how {\sc clr} leaves the spectrum clean of the interference and
keeps the intrinsic detector noise.

For the non-harmonic lines, after examining them in detail and rejecting 
the possibility of the beats, we assume a noninteger harmonic of the
supply
\be
y(t)= \alpha(t) M(t)^q \ ,
\ee
where $\alpha(t)$ is a slowly varying complex amplitude,
$M(t)$ is a reference wave form corresponding the fundamental harmonic
(that can be obtained using {\sc clr}), but $q$ is a real number, not
only an integer as in the case of the harmonics (and it can also
drift in time $q(t)=q_0+\delta q(t)$, where $\delta q(t)$ is small in
comparison to $q_0$). See [4] for details.
The justification for this model is simply the success 
we have  in removing the interference as we show
in figures 3 and 4.

%
%
\begin{figure}[h] 
\begin{center}
\centerline{\vbox{ 
\psfig{figure=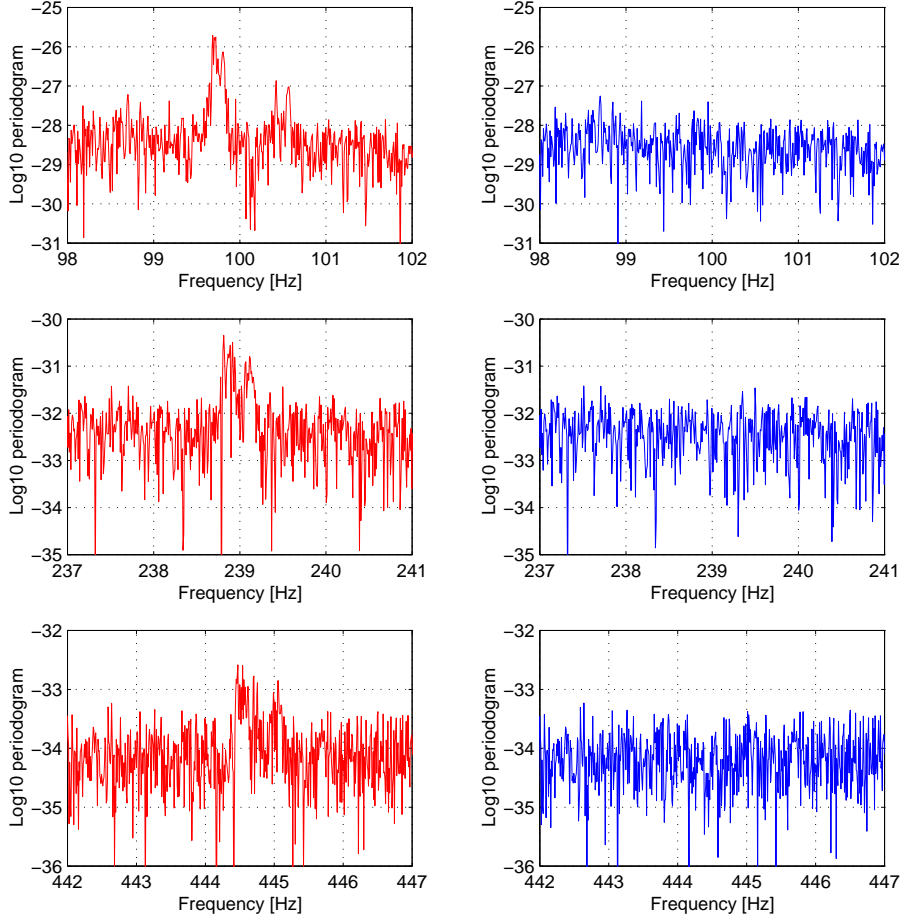,width=12.0cm} 
}} 
\end{center}
\caption[]{Decimal logarithm of the periodogram 
of $2^{19}$ points  of the prototype data. 
 (Left) Details of the lines
near 99.5, 100.5, 239 and 445 Hz. (Right) The same data after removing
the electrical interference.
} 
\end{figure} 
%

%
\begin{figure}[h]
\begin{center}
\begin{tabular}{@{}l|r@{}}
\psfig{file=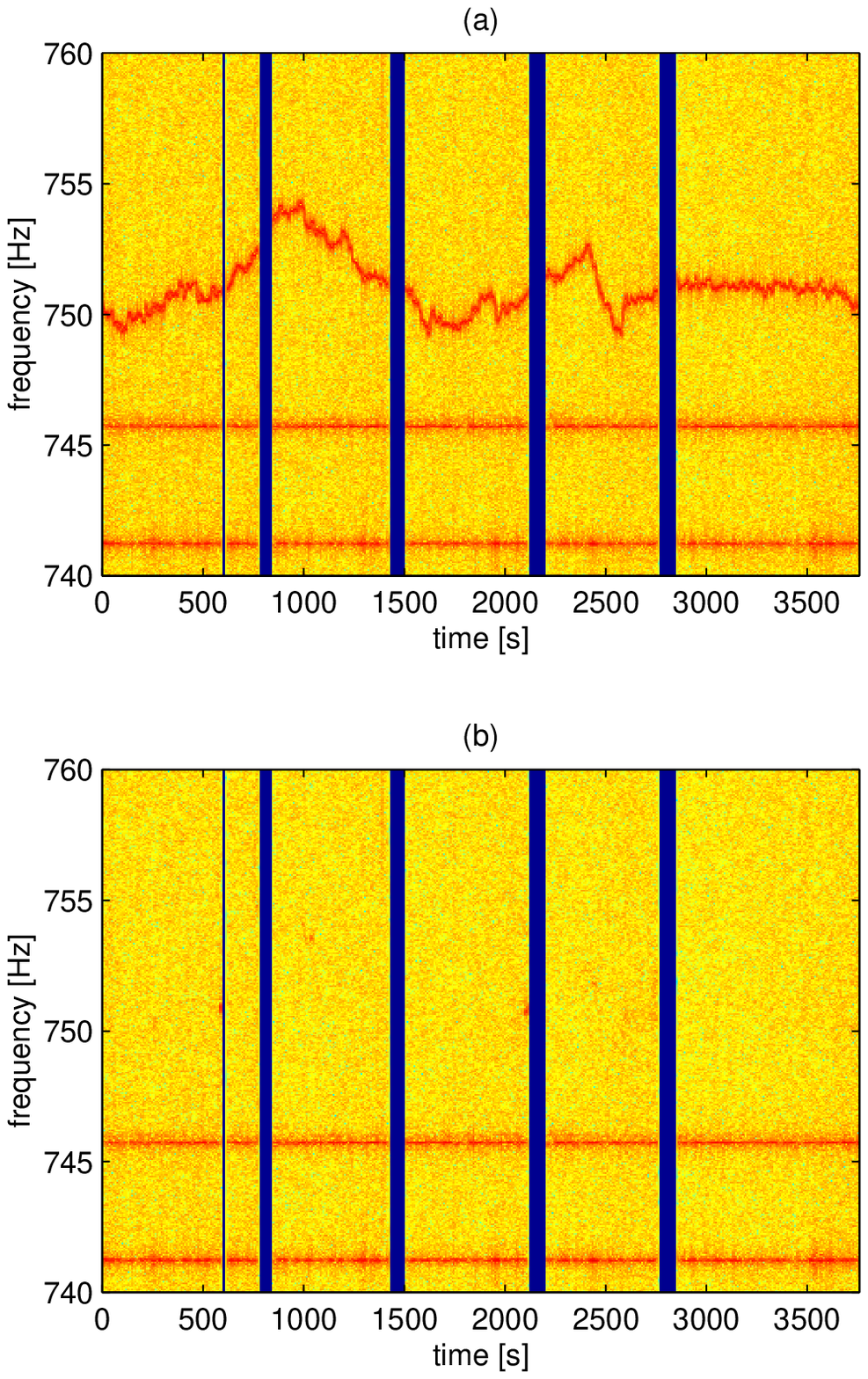,width=60mm,clip=} &
\psfig{file=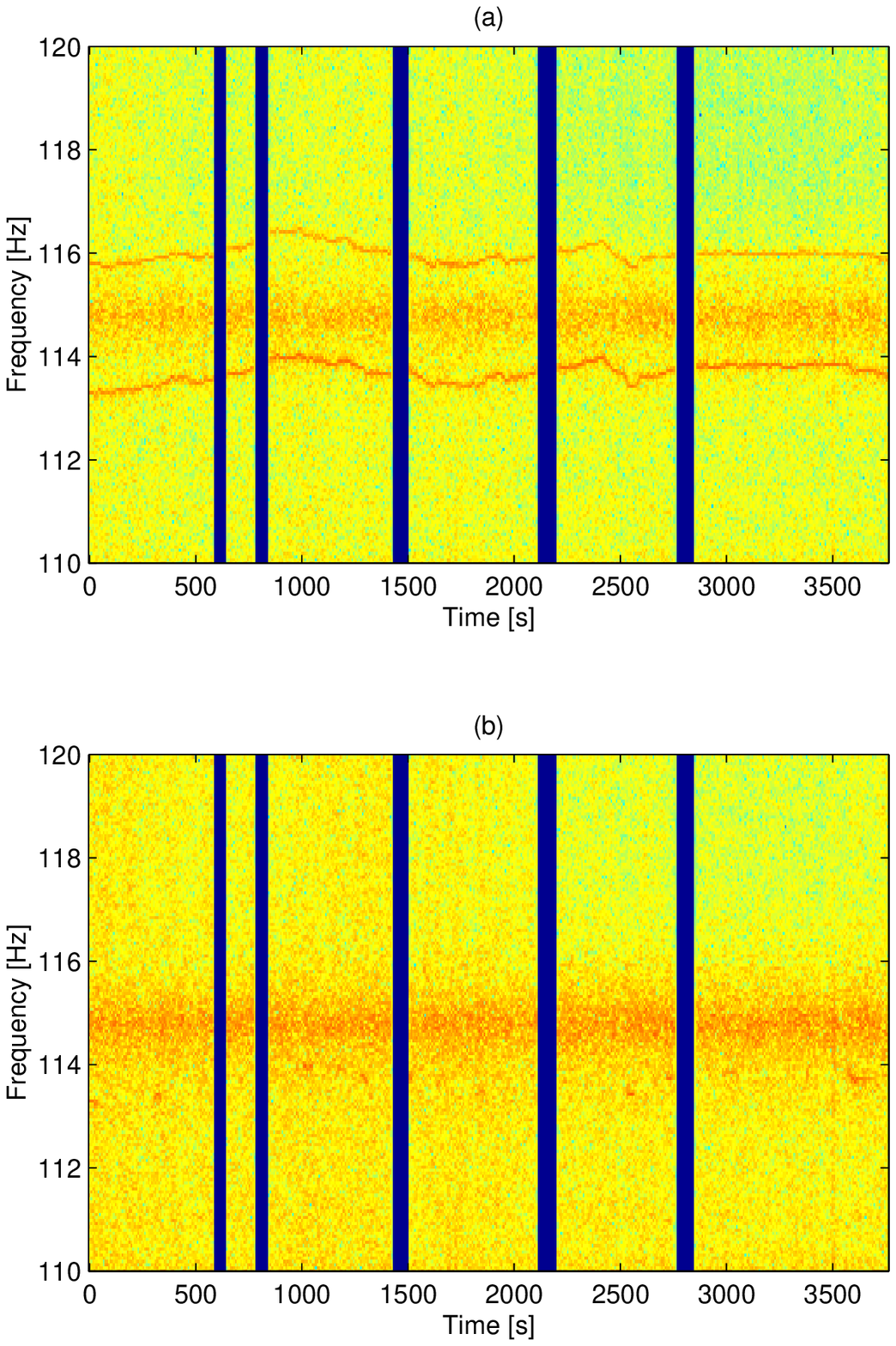,width=63mm,clip=} 
\end{tabular}
\end{center}
\caption[T]{  Comparison of a zoom of the spectrogram. 
 (Left) (a) is obtained from the prototype data. We can observe
the wandering of the incoming electrical 
signal. The other two  remaining lines
at constant frequency correspond to violin modes. (b) The same spectrogram
as in (a) after applying {\sc clr}, showing how the 
electrical interference is 
completely removed.
(Right) 
(a) The prototype data with two anomalous
lines near 114 and 116 Hz. (b) The same spectrogram as in (a)
after removing the interference.
}
\label{all}
\end{figure}

In figure 4,  we compare zooms of the spectrogram. There we can see the 
performance of both algorithms  on the whole data stream, showing how
the lines due to electrical interference in the initial data are removed.


We are interested in studying possible side effects of the line removal
on the statistics of the noise in the time domain.  For the Glasgow data, 
the standard deviation value is around
1.50 Volts. After the line removal, the standard deviation is reduced to
 1.05 Volts. This indicates that a huge amount
of power has been removed. 
Further analysis reveals that values of
skewness and kurtosis are getting closer to zero after the line removal.
Values of skewness and kurtosis 
near  zero suggest a Gaussian nature. Therefore, we are
interested in studying the possible reduction of the level of non-Gaussian 
noise. To this end, we take a piece of data and we study
their histogram, calculating the number of events that lie between
different equal intervals. If we plot the logarithm of the 
number of events versus  $(x-\mu)^2$, where $x$ is the central position of
the interval and $\mu$ is the mean, in case of a Gaussian distribution,
all points should fit on a straight line of slope $-1/2\sigma^2$, where
$\sigma$ is the standard deviation. We observe that this is not the
case. See figure 5. Although, both distributions  seem to have
a linear regime, they present a break and then a very heavy tail.
The two distributions are very different. This is mainly due to the
change of the standard deviation.
We can zoom the \lq linear' regime and change the scale in 
the abscissa  to $(x-\mu)^2/(2\sigma^2)$. Then, any Gaussian
distribution should fit into a straight line of slope -1. 
We observe that after removing the interference, it follows a 
Gaussian distribution quite well up to $4\sigma$. 
The original Glasgow data does not fit   a straight line anywhere.

\begin{figure}[h!]
\begin{center}
\centerline{\vbox{ 
\psfig{figure=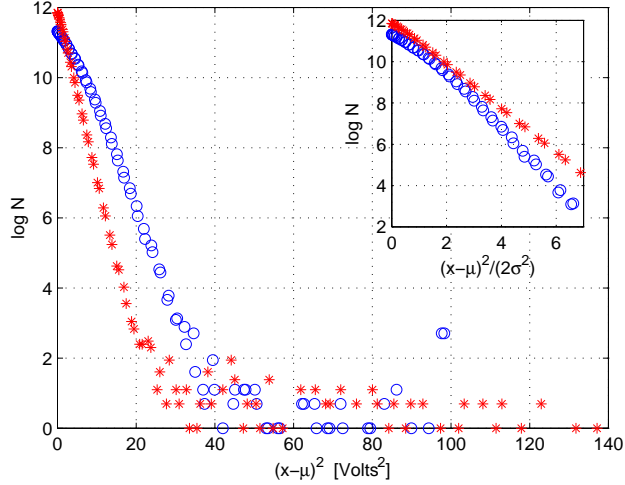,width=9.0cm} 
}} 
\end{center}
\caption[]{ Comparison of the logarithm  plot of the histogram 
for $3.2\times 2^{19}$ points 
as a function of 
$(x-\mu)^2$. The circles correspond to the Glasgow data
and the stars to the same data after removing the electrical interference.
In the right-hand corner, there is zoom  of the original figure, but
rescaled so that
the abscissa corresponds to $(x-\mu)^2/2\sigma^2$.
} 
\end{figure} 
%

We have also applied
 two statistical tests to the
data: the chi-square test  that measures  the discrepancies
between binned distributions,  and the one-dimensional Kolmogorov-Smirnov
test  that measures the differences between  cumulative 
distributions of a continuous data.
In both tests,  the significance  probability increased
after removing the electrical interference, showing  that these
procedures suppress some non-Gaussian noise, although, generally 
speaking, the distribution was still 
non-Gaussian in character. See [3] for details.

\section{References}
\re
1.\  Abramovici A. et al.\ 1996,  Phys. Lett. A 218, 157
\re
2.\ Robertson D.I. et al.\  1995, Rev. Sci. Instrum. 66, 4447
\re
3.\ Sintes A. M., Schutz B.F.\ 1998, Phys. Rev. D, 58, 122003
\re
4.\ Sintes A. M., Schutz B.F.\ 1999, Phys. Rev. D, 60, 062001

%

\end{document}